\documentstyle[12pt]{article}
\newcommand{\newsection}{    % Numeration of eqs. is automatic
\setcounter{equation}{0}
\section}
\newcommand{\tr}[1]{\,{\rm tr}\,#1}

\def\e{{\,\rm e}\,}

\def\eop{\vspace*{\fill}\pagebreak}
\def\be{\begin{equation}}
\def\ee{\end{equation}}
\def\bea{\begin{eqnarray}}
\def\eea{\end{eqnarray}}
\def\LA{\left\langle}
\def\RA{\right\rangle}
\newcommand{\rf}[1]{(\ref{#1})}
\newcommand{\eq}[1]{Eq.~(\ref{#1})}

\def\a{\alpha}
\def\b{\beta}

\def\l{\lambda}
\def\ka{\kappa}
\def\om{\omega}
\newcommand{\ie}{{\it i.e.}\ }
\newcommand{\p}{{\prime}}
\newcommand{\ra}{\rightarrow}
\newcommand{\fr}[2]{{\textstyle {#1 \over #2}}}
\newcommand{\non}{\nonumber \\*}

\def\ga{\mathrel{\mathpalette\fun >}}
\def\fun#1#2{\lower3.6pt\vbox{\baselineskip0pt\lineskip.9pt
\ialign{$\mathsurround=0pt#1\hfil##\hfil$\crcr#2\crcr\sim\crcr}}}
\begin{document}

\begin{titlepage}
\begin{flushright}
YM-5-94 \\
August, 1994 \\
\small{hep-ph/9408337}
\end{flushright}
\vspace*{.1cm}

\begin{center}
{\LARGE Threshold Multiparticle Amplitudes} \\
\vspace{0.2cm}{\LARGE in $\phi^4$ Theories at Large $N$}
\end{center} \vspace{.3cm}

\begin{center}
{\large Yu.\ Makeenko}\footnote{~E--mail:
\ makeenko@vxitep.itep.msk.su \ / \
 makeenko@nbivax.nbi.dk \ }
\vskip 0.2 cm
%\\ \mbox{} \\
{\it Institute of Theoretical and Experimental Physics,}
\\ {\it B. Cheremushkinskaya 25, 117259 Moscow, Russian Federation}
\\ \vskip .1 cm
and  \\  \vskip .1 cm
{\it The Niels Bohr Institute,} \\
{\it Blegdamsvej 17, 2100 Copenhagen, Denmark}
\end{center}

\vskip 0.3 cm
\begin{abstract}
%\noindent
I review some recent work on the problem of multiparticle production
in a $\phi^4$-theory with an $O(N)$ symmetry.
Threshold amplitudes with fixed number of produced particles
are exactly calculated at large-$N$ to all loops
and vanish on mass shell for $2$$\ra$$n$ when $n$$>$$2$
due to a dynamical symmetry.
I consider an extension to the cases when the $O(N)$ symmetry is softly
broken by masses including spontaneous breaking of
a remaining reflection symmetry.
The exact solutions are obtained by the Gelfand--Diki\u{\i} technique
of finding the diagonal resolvent of the Schr\"{o}dinger operator
which emerges due to factorization at large $N$.
I report also some new results on the diagonal
resolvent for a general P\"{o}schl--Teller
potential, which could be useful for calculations of multiparticle
amplitudes in the standard model.
\end{abstract}

\vspace{1cm}
\noindent
Talk at the International Seminar {\sl Quarks--94\/}, Vladimir, Russia,
May 12--17, 1994.%
\footnote{~Similar talks are given at {\sl 1st annual meeting on high energy
phenomenology\/}, Athens, Greece, June 9--11, 1994 and
{\sl XVIII-th triangular meeting on modern field theory}, Copenhagen,
Denmark, June 27--29, 1994.}

\eop
\end{titlepage}
\setcounter{page}{2}

\newsection{Introduction and r\'{e}sum\'{e}}

\subsection{$n!$}

The history of the subject
began in 1990 when Ringwald~\cite{Rin90} and Espinosa~\cite{Esp90}
utilized $n!$, which is due to $n$ identical scalar particles in a
final state, to get a large amplitude of B+L violating processes
in the standard model due to instantons in the dilute
gas approximation.
Soon after Cornwall~\cite{Cor90} and
Goldberg~\cite{Gol90} applied a similar idea to a problem
of perturbative multiparticle production in the $\phi^4$ theory with
a small coupling constant $\l$.
Some other early references include also~\cite{Vol91,CT92,GV91}.

In the $\phi^4$ theory one considers
the process when a virtual particle with the $4$-mo\-men\-tum $(nm,\vec{0})$
produces $n$ on-mass-shell particles of mass $m$ at rest with
the $4$-mo\-men\-tum
$(m,\vec{0})$ each. Such a process may appear in a high energy collision of
quark and antiquark when $n$ Higgs particles are produced. Since the amplitude
of the process in proportional to $n!$:
\be
a(n) \propto n!\,,
\ee
a naive estimate of the cross-section in the tree approximation is
\be
  \sigma_{q\bar{q}\ra nH} \sim \frac{1}{n!} |a(n)|^2 \sim n!\, \l^{n-1} \,,
\label{estimate}
\ee
which may become large for $n\ga 1/\l$.

A special comment is needed about ``triviality'' of the $\phi^4$ theory.
If the $\phi^4$ theory were a fundamental one,
the renormalized value of $\l$ would vanish. A modern look at this problem
is that it is no longer fundamental at the distances
\be
r \sim m^{-1} \e^{-8\pi^2/3\l}
\ee
which are very small if (renormalized) $\l$ is small
and play a role of a cutoff. Therefore, the estimate~\rf{estimate}
makes sense for small enough $\l$ when $n\ll\exp{(8\pi^2/3\l)}$ can still be
large.

\subsection{The tree level}

The explicit results
for the tree amplitudes at threshold in the $\phi^4$ theory
\be
a(n) = - (n^2-1)\, n!\, m^2 \left( \frac{\l}{c\,m^2} \right)^{(n-1)/2},
\label{tree}
\ee
where $c=8$~\cite{Vol92} for an unbroken reflection symmetry $\phi\ra-\phi$
and $c=2$~\cite{AKP93a} for a broken one,
demonstrate the expected factorial growth.
These results are obtained solving exact tree-level recurrence
relations between the amplitudes $a(n)$ with different $n$,
which is an alternative to semiclassical evaluations
at large $n$.
A bridge between the two approaches relies on a
functional technique~\cite{Bro92} which turns out to be very useful to go
beyond the tree level.

\subsection{Unitarity violation}

While the threshold amplitudes
have a vanishing phase volume, they are a lower bound for tree
amplitudes with nonvanishing spatial momenta of produced particles~\cite{CT92}.
This bound looks like the r.h.s.\ of \eq{tree} with $m$ substituted by
the variable $\om=E-(n-1)m$ where $E$ is the total energy of produced
particles in the rest frame ($\om=m$ if all particles are produced
at rest) and $c=9$ or $c=8/3$~\cite{AKP92b,Vol92b} for the unbroken or broken
symmetry, respectively.

This behavior of the tree amplitudes leads to a violation of
unitarity~\cite{CT92,AKP92b,Vol92b,AKP92c}. Moreover, if one reconstructs
the perturbative expansion by the unitarity relation, it
means under some natural assumptions that the perturbation theory
breaks down starting from some finite order.
Therefore, loops should be taken into account.

\subsection{The nullification}

The one-loop calculation of the $1$$\ra$$n$ amplitude
by Voloshin~\cite{Vol93a} discovered
a very interesting property of the tree $2$$\ra$$n$ threshold amplitude which
vanishes for $n$$>$$4$ when incoming particles are on mass shell.
This nullification occurs for dynamical reasons because of the cancellation
between diagrams.
It has been extended~[15--25] %%%%\cite{Smi93}--\cite{Smi94}
to more general models and is crucial for
calculations of the amplitude $1$$\ra$$n$ at the one-loop level.
A dynamical symmetry which may be responsible for the nullification has been
discussed in Ref.~\cite{LRT93b}.
An interesting question is
which properties of the tree and one-loop amplitudes could survive
in the full theory which includes all loop diagrams.

\subsection{$O(\infty)$ symmetry}

The simplest exactly solvable to all loops
model is the $O(N)$-symmetric $\phi^4$ theory
in the large-$N$ limit when only bubble
diagrams contribute if there is no multiparticle production.
The amplitudes of multiparticle production at threshold
are calculated for this model in the large-$N$ limit at fixed number
of produced particles $n$ in Ref.~\cite{Mak94}.
The results, which are reviewed in Sect.~2 below,
are quite similar to the tree amplitudes while the effect of
loops is the renormalization of the coupling constant and mass.
The form of the solution is due to the fact that the exact amplitude
of the process $2$$\ra$$n$ vanishes on mass shell for $n$$>$$2$ at large $N$
when averaged over the $O(N)$ indices of incoming particles.
An extension~\cite{AJM94} to the cases when the $O(N)$ symmetry is softly
broken by masses including spontaneous breaking of
a remaining reflection symmetry is considered in Sect.~3.

\subsection{The rescattering problem}

Since the large-$N$ multiparticle amplitudes are similar to the tree-level
ones, they are not very instructive from the viewpoint of unitarity
restoration by loops (while agree with unitarity by themselves).
More essential for this are loops associated
with rescatterings of the produced particles which are
controlled at large $n$ by the parameter $n^2 \l$~\cite{Vol93a,Smi93}.
As $\l\sim1/N$ they give
corrections to the large-$N$ amplitudes which are controlled by
$n^2 /N$.

The rescattering contribution is expected to be calculable by semiclassical
methods~\cite{Vol93c,GV93}. Nice explicit results showing how unitarity
is restored by this mechanism
are obtained~\cite{RS94} in $2+1$ dimensions by an alternative method
of summing infrared logarithms with the aid of renormalization group.
The conjecture about exponentiation of the $n^2 \l$ (or $n^2/N$)
corrections, which is confirmed by various considerations~\cite{LRST94},
can help to solve the problem.

\subsection{Yet other applications?}

The techniques developed in studying multiparticle production
might be useful for other problems (some of them are mentioned in Sect.~4).
An idea is that for the given kinematics, which reduces the system
at the tree and one-loop levels or at large $N$ to an integrable
lower-dimensional one, calculations could be simpler than, say, for
all momenta to be of the same order.

Equations of this type appear in many problems with an emission of energy.
In particular, the large-$N$ Schwinger--Dyson equations considered
in Subsect.~2.2 coincide with those~\cite{Mot94} describing
a non-equilibrium time evolution of quantum systems.%
\footnote{~I thank E.~Mottola for this comment.}
The difference is that the latter problem is in real
rather than in imaginary time and, what is more essential, the boundary
conditions are quite different.

\newsection{Unbroken $O(\infty)$ symmetry [27]}

\subsection{The definitions}

The $O(N)$-symmetric $\phi^4$ theory is defined in the
$4$-dimensional Minkowskian space by the Lagrangian
\be
{\cal L} = \fr 12 (\partial_\mu \phi^b) (\partial_\mu \phi^b)
-\fr 12 m^2({\phi}^b{\phi}^b)
-\fr 14 \l ({\phi}^b{\phi}^b)^2 \,,
\label{lagrangian}
\ee
where $b=1,\ldots,N$ and the summation over repeated indices is implied.

The amplitudes $a^b_{b_1\cdots b_n}(n)$ of production of $n$
on-mass-shell particles with the $O(N)$ indices
${b_1\ldots b_n}$ at rest by a (virtual) particle
with the $O(N)$ index $b$ and the energy $nm$ are given by
the generating function~\cite{AKP93a,Bro92}
\bea
\Phi^b(\tau) &=& \sum_{n\geq1} \frac{1}{n!} \LA b_1\ldots b_n |
\phi^b(\vec{0},-i\tau) |0 \RA \xi^{b_1} \cdots \xi^{b_n} \non
 &=& m \xi^b \e^{m\tau}  + \sum_{n\geq3}
 \frac{a^b(n)}{n!(n^2-1)} \e^{nm\tau} m^{n-2},
\label{Phi}
\eea
where $\tau= i x_0$ is the imaginary-time variable and
\be
a^b(n) = a^b_{b_1\cdots b_n}(n)\, \xi^{b_1}\ldots \xi^{b_n} \,.
\label{defa}
\ee
The multiplication by
the $O(N)$ vectors $\xi^{b_i}$ results in a symmetrization over the
$O(N)$ indices of the produced particles.
The tree-level estimate gives
\be
a^b(n) \sim \left(\l{\xi}^2\right)^{\frac{n-1}{2}} \xi^b \,.
\label{order}
\ee
Since $\l \sim 1/N$ in the large-$N$ limit, we choose ${\xi}^2 \sim N$
for all the amplitudes~\rf{order} to be of the same order in $1/N$.

It is convenient to introduce the coherent state~\cite{AJM94}
\be
\LA  \Xi | \right. ~~=
\sum_{n=0}^\infty \frac{1}{n!}
\xi^{b_1} \cdots \xi^{b_n}\,
\LA  b_1\ldots b_n | \right. ~~\,.
\label{coherent}
\ee
Then \eq{Phi} can be written as
\be
\Phi^b(\tau) = \LA \Xi | \phi^b(\vec{0},-i\tau) |0 \RA \,.
\label{PhiX}
\ee

The second quantity which appears in the set of the Schwinger--Dyson
equations of the next subsection is the amplitude
$D^{ab}_{b_1\cdots b_n}(n;p)$ of the process
when two particles $a$ and $b$ with the $4$-momenta
$p+nq$ and $-p$, respectively, produce $n$ on-mass-shell particles
with the $O(N)$ indices ${b_1\ldots b_n}$
and the equal $4$-momenta $q=(m,0)$ in the rest frame. We define again
\be
D^{ab}(n;p)=D^{ab}_{b_1\cdots b_n}(n;p)\, \xi^{b_1}\ldots \xi^{b_n}
\label{defD}
\ee
and introduce the generating function
\bea
G^{ab}_\om(\tau,\tau^\p) &=& \int \frac{d^3 \vec{x}}{(2\pi)^3}
\e^{i\vec{p}\vec{x}}
\LA \Xi | \phi^a(\vec{x},-i\tau) \phi^b(\vec{0},-i\tau^\p)| 0 \RA
- \delta^{(3)}(\vec{p}) \Phi^a(\tau) \Phi^b(\tau^\p) \non
&=&\frac{\delta^{ab}}{2\om} \e^{-\om|\tau-\tau^\p|} +
 \int \frac{d\epsilon}{2\pi} \sum_{n=0}^\infty
\e^{(\epsilon+nm)\tau-\epsilon\tau^\p}\; D^{ab}(n;p) \,.
\label{Dphi}
\eea
Here
\be
\om =\sqrt{\vec{p}\,{}^2+m^2}~\,,
\ee
$p=(\epsilon, \vec{p})$ and the subtraction cancels disconnected parts
which are not included in the definition of $D^{ab}(n;p)$.
The meaning of $G^{ab}_\om(\tau,\tau^\p)$ is that of the Green function in the
imaginary-time representation while $D^{ab}(n;p)$ is the one in the energy
representation.

\subsection{Scwinger--Dyson equations}

The Schwinger--Dyson equations can be
extracted from the following identity
\be
\LA \Xi \left| \frac{\delta S[\phi]}{\delta \phi^a(x)}
F[\phi] \right|0\RA = i
\LA \Xi \left| \frac{\delta F[\phi]}{\delta \phi^a(x)}
\right|0\RA
\label{identity}
\ee
where $S$ is the action of the model and $F[\phi]$ is an
arbitrary functional of $\phi$.
\eq{identity} results from the invariance of the
measure in the path integral
under an arbitrary variation of $\phi^a(x)$.
Substituting $F=1$, one gets from \eq{identity}
\be
\left\{ \frac{\partial^2}{\partial \tau^2}+
\frac{\partial ^2}{\partial \vec{x}^2} -m^2\right\}
\LA \Xi |\phi^a(\vec{x},-i\tau) |0 \RA
- \l \LA \Xi| \phi^2(\vec{x},-i\tau)  \phi^a(\vec{x},-i\tau) | 0 \RA =0 \,.
\label{opX}
\ee

In the large-$N$ limit one can split the matrix elements of
the operator
$\phi^2 \phi^a $  as
\be
\LA b_1\ldots b_n | \phi^2 \phi^a  |0\RA
=\sum_{p,p'} \sum_{n_1 +n_2 =n} \frac{n!}{n_1!n_2!}
\LA p(b_1)\ldots p(b_{n_1})  | \phi^2 |0\RA
 \LA p(b_{n_1 + 1})\ldots p(b_n) |\phi^a |0\RA
\label{factorization}
\ee
where $p$ and $p^\prime$ stand for permutations.
This formula holds for $n$ finite as $N\ra\infty$ and
extends  the standard {\it factorization\/} of $O(N)$-singlet operators
at large $N$ to the case of multiparticle production.
According to the definition~\rf{coherent},
we rewrite \eq{factorization} in the form
\be
\LA \Xi| \phi^2(\vec{x},-i\tau)  \phi^a(\vec{x},-i\tau) | 0 \RA
=\LA \Xi | \phi^2(\vec{x},-i\tau)  | 0 \RA
\LA \Xi | \phi^a(\vec{x},-i\tau) | 0 \RA \,.
\label{fact}
\ee

Using Eqs.~\rf{PhiX}, \rf{fact} and translational invariance, \eq{opX}
can be written as
\be
\left\{\frac{d^2}{d\tau^2} -m^2- v(\tau)\right \} \Phi^a(\tau) =0\,,
\label{eq1}
\ee
where we have denoted
\be
v(\tau)\equiv \l \LA \Xi| \phi^2(\vec{0},-i\tau) | 0 \RA =
 \l \Phi^2(\tau) +
 \frac{\l}{2\pi^2}\int_{m^2}d\om \;\om \sqrt{\om^2-m^2} \;
G^{aa}_\om(\tau,\tau)
\label{uvsG}
\ee
and the last equality holds due to the definition~\rf{Dphi}.

The meaning of the amplitude $G^{aa}$ which is
summed over the $O(N)$ indices of incoming particles
becomes clear after the obvious decomposition
\be
G^{ab}_\om(\tau,\tau^\p) =\left(\delta^{ab}
-\frac{\xi^a\xi^b}{\xi^2}\right)
G^T_\om(\tau,\tau^\p)+ \frac{\xi^a\xi^b}{\xi^2}
G^{S}_\om(\tau,\tau^\p)\,.
\label{structure}
\ee
At large $N$ one gets
\be
\frac{1}{N} G^{aa}_\om(\tau,\tau^\p) =
\left(1- \frac{1}{N}\right)G^T_\om(\tau,\tau^\p) +
\frac 1N G^{S}_\om(\tau,\tau^\p) \ra G^T_\om(\tau,\tau^\p)\,.
\ee
Only $G^T$ enters the Schwinger--Dyson equations at large $N$.

In order to close the set of equations, one derives an equation for
the propagator $G^{T}_\om(\tau,\tau^\prime)$ which corresponds to
$F=\phi^a(\vec{0},-i\tau^\p)$ in \eq{identity}. Using the large-$N$
factorization and integrating over $d^3 \vec{x}$, one gets
\be
\left\{\frac{d^2}{d\tau^2} -\om^2- v(\tau) \right\}
 G^{T}_\om(\tau,\tau^\p)=
-\delta(\tau-\tau^\p) \,.
\label{eqforG}
\ee
The set of Eqs.~\rf{eq1}, \rf{uvsG} and \rf{eqforG} is closed and
unambiguously determined the solution presented in the next subsection.

\subsection{The exact solution}

Eqs.~\rf{eq1}, \rf{uvsG} and \rf{eqforG} look like a quantum mechanical
problem which can be solved~\cite{Mak94} applying the Gelfand--Diki\u{\i}
technique~\cite{GD75} which is described in Subsect.~\ref{gd}.
The solution, which satisfies proper boundary conditions as $\tau\ra-\infty$,
reads
\be
 G^{T}_\om(\tau,\tau)
 = \frac{1}{2\om} -\frac{\bar{\l} \Phi^2(\tau)}{4\om(\om^2-m_R^2)} \,,
\label{exact}
\ee
\be
v_R(\tau) = \bar{\l} \Phi^2(\tau)\,,
\label{exact1}
\ee
where
\be
\bar{\l}= \frac{\l_R}{1+\frac{\l_R N}{8\pi^2}} \,,
\label{lbar}
\ee
and
\be
\Phi^a(\tau) =
 \frac{\xi^am_R \e^{m_R\tau}}
{1- \frac{\bar{\l} \xi^2}{8}\e^{2m_R\tau}} \,.
\label{finalPhi}
\ee

The solution is expressed via
the renormalized mass $m_R$ and coupling $\l_R $ which are
related to the bare ones $m$ and $\l$ by
\be
m^2=m^2_R -
\frac{\l N}{4\pi^2}\int_{m_R^2}d\om \; \sqrt{\om^2-m^2_R}
\label{mR}
\ee
and
\be
\frac{1}{\l} = \frac{1}{\l_R}- \frac{N}{8\pi^2}\int_{m_R^2} d\om\,
 \frac {\sqrt{\om^2-m^2_R}} {\om^2} \,,
\label{lR}
\ee
which exactly coincide with the standard
renormalizations of mass and the coupling constant at large $N$.
The renormalization of $v(\tau)$ is given by
\be
v_R(\tau) = v(\tau) +m^2-m^2_R \,.
\label{uR}
\ee

Eqs.~\rf{exact}, \rf{finalPhi} coincide with
the large-$N$ limit of the tree-level results for
$G^{T}_\om(\tau,\tau)$ \cite{Smi93a} and
$\Phi^a(\tau)$ \cite{Bro92} in $O(N)$-symmetric $\phi^4$ theory,
while the only effect of loops, besides the renormalization, is
the change of $\l_R$ given by \eq{lbar}.

\begin{figure}[tbp]
\unitlength=1.00mm
\linethickness{0.6pt}
\centering
\begin{picture}(118.00,68.00)(15,70)
\put(23.00,78.00){\makebox(0,0)[cc]{{\Large a)}}}
\put(5.00,128.00){\line(1,0){18.00}}
\put(5.00,100.00){\line(1,0){18.00}}
\put(23.00,100.00){\line(0,1){28.00}}
\put(24.50,128.00){\line(2,1){12.00}}
\put(24.50,100.00){\line(2,-1){12.00}}
\put(24.50,128.00){\line(2,-1){12.00}}
\put(24.50,100.00){\line(2,1){12.00}}
%2
\put(70.00,78.00){\makebox(0,0)[cc]{{\Large b)}}}
\put(70.00,114.00){\line(-2,1){18.00}}
\put(70.00,114.00){\line(-2,-1){18.00}}
\put(71.50,114.00){\line(2,1){18.00}}
\put(71.50,114.00){\line(1,-1){18.00}}
\put(81.50,105.50){\line(4,-1){8.00}}
\put(81.50,105.50){\line(2,1){8.00}}
%3
\put(125.00,114.00){\line(-2,1){18.00}}
\put(125.00,114.00){\line(-2,-1){18.00}}
\put(126.50,114.00){\line(2,-1){18.00}}
\put(126.50,114.00){\line(1,1){18.00}}
\put(125.00,78.00){\makebox(0,0)[cc]{{\Large c)}}}
\put(136.50,122.50){\line(4,1){8.00}}
\put(136.50,122.50){\line(2,-1){8.00}}
\end{picture}
\caption[x]   {\hspace{0.2cm}\parbox[t]{13.5cm}
{\small
   The tree diagrams for $D^{T}(4;p)$ at large $N$. The continuous lines
   are associated with propagation of $O(N)$ indices.
   The contributions of the diagrams a), b) and c) equal, respectively,
   $-1/4$, $1/8$ and $1/8$ for the given on-mass-shell kinematics
   when the momenta of the incoming particles (in the units of mass)
   are $(2,\sqrt{3},0,0)$ and $(2,-\sqrt{3},0,0)$,
   while that of each produced particle is $(1,0,0,0)$.
   The sum of three diagrams vanishes which illustrates the nullification
   of $2$$\ra$$4$ on mass shell at the tree level.
   }}
   \label{fig5}
   \end{figure}
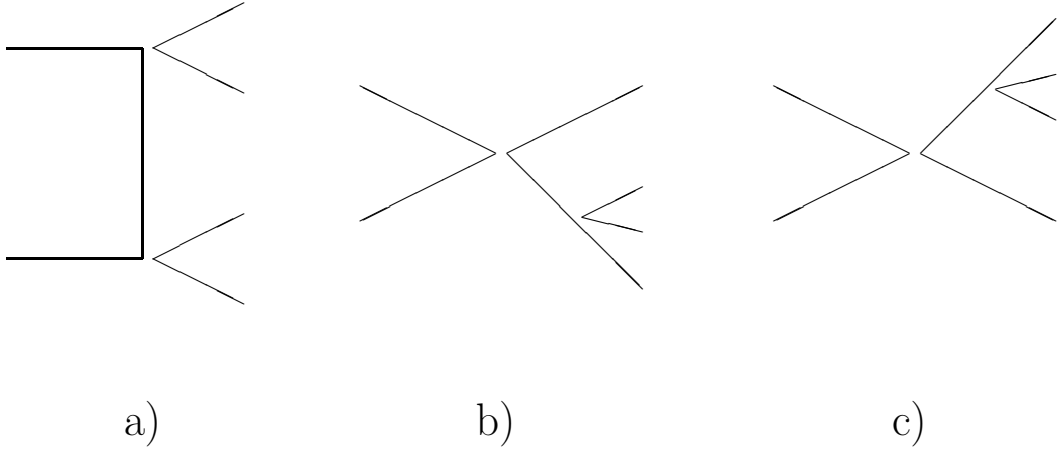

The fact that~\rf{exact}
has a pole only at $\om^2=m_R^2$
means the nullification of the
on-mass-shell amplitudes $D^{T}(n;p)$ for $n$$>$$2$.
An analogous property holds~\cite{AKP93b}
at $N=1$ for the tree level amplitudes
in the case of the sine-Gordon potential and for the
$\phi^4$-theory with spontaneously broken symmetry~\cite{Smi93}.
It can be explicitly demonstrated by the cancellation of
the tree diagrams for the process $2$$\ra$$4$ which
are depicted in Fig.~\ref{fig5}.

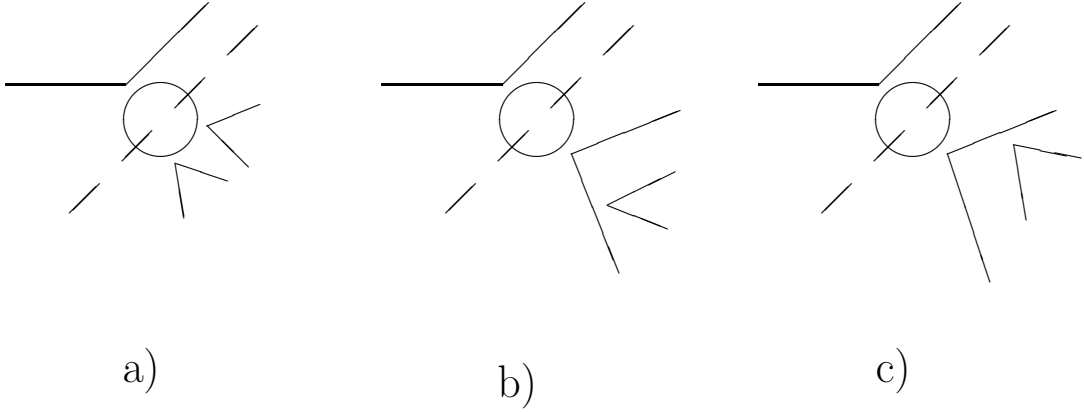
\begin{figure}[t]
%\begin{figure}[tbp]
\unitlength=1.00mm
\linethickness{0.6pt}
\centering
\begin{picture}(118.00,60.00)(19,83)
%\begin{picture}(147.00,139.00)
\put(23.00,90.00){\makebox(0,0)[cc]{{\Large a)}}}
\put(5.00,128.00){\line(1,0){16.00}}
\put(21.00,128.00){\line(1,1){11.00}}
\put(25.60,123.40){\circle{10.00}}
\put(31.80,122.50){\line(5,2){7.00}}
\put(31.80,122.50){\line(1,-1){5.50}}
\put(27.50,117.50){\line(3,-1){7.00}}
\put(27.50,117.50){\line(1,-6){1.20}}
\put(73.00,88.00){\makebox(0,0)[cc]{{\Large b)}}}
\put(55.00,128.00){\line(1,0){16.00}}
\put(71.00,128.00){\line(1,1){11.00}}
\put(75.60,123.40){\circle{10.00}}
\put(80.20,118.80){\line(5,2){14.50}}
\put(80.20,118.80){\line(2,-5){6.33}}
\put(85.00,112.00){\line(5,-2){8.00}}
\put(85.00,112.00){\line(2,1){9.00}}
\put(123.00,90.00){\makebox(0,0)[cc]{{\Large c)}}}
\put(105.00,128.00){\line(1,0){16.00}}
\put(121.00,128.00){\line(1,1){11.00}}
\put(125.60,123.40){\circle{10.00}}
\put(130.20,118.80){\line(5,2){14.50}}
\put(130.20,118.80){\line(1,-3){5.67}}
\put(139.00,120.00){\line(1,-6){1.67}}
\put(139.00,120.00){\line(5,-1){9.00}}
\thinlines
\multiput(26.10,124.90)(7.0,7.0){2}{
\line(1,1){4.00}}
\multiput(23.10,121.90)(-7.0,-7.0){2}{
\line(-1,-1){4.00}}
\multiput(76.10,124.90)(7.0,7.0){2}{
\line(1,1){4.00}}
\multiput(73.10,121.90)(-7.0,-7.0){2}{
\line(-1,-1){4.00}}
\multiput(126.10,124.90)(7.0,7.0){2}{
\line(1,1){4.00}}
\multiput(123.10,121.90)(-7.0,-7.0){2}{
\line(-1,-1){4.00}}
\end{picture}
\caption[x]   {\hspace{0.2cm}\parbox[t]{13.5cm}
{\small
   The one-loop diagrams for $a^{b}(5;p)$ at large $N$.
   The dashed lines represent unitary cuts over two-particle intermediate
   states.
   The imaginary part vanishes after summing over all three diagrams
   since the amplitude $D^{T}(4;p)$ vanish on-mass-shell at the tree level
   as is illustrated by Fig.~\ref{fig5}.
   }}
\label{real}
\end{figure}

The large-$N$
amplitude~\rf{finalPhi} is {\it real\/} which is due to the
nullification. This is different from the
$N=1$ one-loop result~\cite{Vol93a} while
is similar to the case of spontaneously broken reflection
symmetry where the $1$$\ra$$n$ amplitude
is real to all orders~\cite{Vol93c}.
The fact that the amplitude $1$$\ra$$5$ is real at the one-loop level
is illustrated by Fig.~\ref{real}.
While the sum of the imaginary parts vanishes, the real part is described by
Eqs.~\rf{lbar}, \rf{finalPhi}.

Analogously, the imaginary parts of the one-loop diagrams for the
$2$$\ra$$4$ on-mass-shell amplitude $D^{T}(4;p)$ are mutually
cancelled for the same reason while the real part is given by Eqs.~\rf{exact},
\rf{lbar} and \rf{finalPhi}. This amplitude is also real to all loops.
Therefore, both amplitudes are real due to the nullification which, in turn,
is due to a hidden dynamical symmetry.

\newsection{Softly broken $O(\infty)$ symmetry~[28]}

\subsection{The $O(N_1)$$\times$$O(N_2)$ model}

The model consists of an $O(N_1)$ multiplet $\chi^b$,
$b=1,...,N_1$, coupled to a $O(N_2)$ scalar field $\phi^{\b}$,
$\b =1,...,N_2$.
The Lagrangian reads
\be
{\cal L} = \fr 12 (\partial_\mu \chi^b) (\partial_\mu \chi^b)+
\fr 12 (\partial_\mu \phi^{\b}) (\partial_\mu \phi^{\b})-
\fr 12 m_1^2({\chi}^b{\chi}^b)
-\fr 12 m_2^2(\phi^{\b} \phi^{\b})
-\fr 14 \l ( {\chi}^b{\chi}^b+\phi^{\b} \phi^{\b})^2\,,
\label{llagrangian}
\ee
where interaction term is $O(N_1$$+$$N_2)$ invariant while the mass
term possesses the $O(N_1) \times O(N_2)$ symmetry for nonequal
masses $m_1$ and $m_2$. All the formulas of this section are
proper extensions of those of Sect.~2 to $m_1\neq m_2$.

We denote by $A^b_{b_1\ldots b_n,\b_1 \ldots \b_k}
(n,k)$ the amplitude of production of $n$
on-mass-shell particles $\chi$ with the $O(N_1)$-indices
${b_1\ldots b_n}$ and of $k$ on-mass-shell particles $\phi$
with the $O(N_2)$-indices $\b_1, \ldots, \b_k$
at rest by a (virtual) particle $\chi$
with the $O(N_1)$-index $b$ and the energy
$nm_1 +km_2$. We also denote by $B^{\b}_{b_1\ldots b_n,\b_1 \ldots \b_k}
(n,k)$ the amplitude of production of $n$
on-mass-shell particles $\chi$ with the $O(N_1)$-indices
${b_1\ldots b_n}$ and of $k$ on-mass-shell particles $\phi$ with
$O(N_2)$-indices $\b_1, \ldots, \b_k$
at rest by a (virtual) particle $\phi$ with the $O(N_2)$-index
$\beta$ and the energy $nm_1 +km_2$. Analogues of \eq{defa} are
\bea
A^b(n,k) = A^b_{b_1\ldots b_n ,\b_1 \ldots \b_k}(n,k)\,
\xi_1^{b_1}\cdots \xi_1^{b_n} \xi_2^{\b_1} \cdots \xi_2^{\b_k} \, ,
\label{defab}\\
B^{\b} (n,k) = B^{\a}_{b_1\ldots b_n ,\b_1 \ldots \b_k}
(n,k)\, \xi_1^{b_1}\cdots \xi_1^{b_n}
\xi_2^{\b_1} \cdots \xi_2^{\b_k} \,.
\label{ddefab}
\eea

An analog of the coherent state~\rf{coherent} is
\be
\LA  \Xi | \right. ~~=
\sum_{n,k=0}^\infty \frac{1}{n!k!}
\xi_1^{b_1} \cdots \xi_1^{b_n}
\xi_2^{\b_1} \cdots \xi_2^{\b_k}\,
\LA  b_1\ldots b_n ,\b_1\ldots \b_k | \right. ~~\,,
\label{ccoherent}
\ee
while the generating functions for the amplitudes~\rf{defab}, \rf{ddefab} are
\be
\Phi^a(\tau) = \LA \Xi | \chi^a(\vec{0},-i\tau) |0 \RA
 = m_1 \xi_1^a \e^{m_1\tau}  + \sum_{n,k\geq3}
\frac{A^a(n,k) \e^{(nm_1+km_2)\tau}}{n!k!((nm_1 +km_2)^2 - m_1^2)}  m_1^n
m_2^k ,
\label{PhiAX}
\ee
\be
\Psi^\a(\tau) = \LA \Xi | \phi^\a(\vec{0},-i\tau) |0 \RA
= m_2 \xi_2^{\a} \e^{m_2\tau}  + \sum_{n,k\geq3}
\frac{B^{\a}(n,k)\e^{(nm_1+km_2)\tau}}{n!k!((nm_1 +km_2)^2 - m_2^2)}
m_2^n m_2^k .
\label{PsiBX}
\ee
The definition of the generating functions $G^{ab}_\chi(\om;\tau,\tau^\p)$
and $G^{\a\b}_\phi(\om;\tau,\tau^\p)$ which extend \rf{defD} and \rf{Dphi}
is obvious.

The Schwinger--Dyson equations which extend Eqs.~\rf{eq1}, \rf{uvsG} and
\rf{eqforG} read
\be
\left\{\frac{d^2}{d\tau^2} -m_1^2- v(\tau)\right \} \Phi^a(\tau) =0\,,
\label{eeq1}
\ee
\be
\left\{\frac{d^2}{d\tau^2} -m_2^2- v(\tau)\right \} \Psi^{\a}(\tau) =0\,,
\label{eqf}
\ee
\be
\left\{\frac{d^2}{d\tau^2} -\om^2- v(\tau) \right\}
 G^{T}_\chi(\om ;\tau,\tau^\p)=
-\delta(\tau-\tau^\p) \,,
\label{eeqforG}
\ee
\be
\left\{\frac{d^2}{d\tau^2} -\om^2- v(\tau) \right\}
G^{T}_{\phi} (\om ;\tau,\tau^\p)=
-\delta(\tau-\tau^\p) \,,
\label{eqforF}
\ee
where
\bea
v(\tau) = \l \LA \Xi | (\chi^2(\vec{0},-i\tau)+\phi^2(\vec{0},-i\tau))
|0\RA =
\l (\Phi^2(\tau)+\Psi^2(\tau)) ~~~~~~~~~~~~~~\non +
\frac{\l N_1}{2\pi^2}\int_{m_1}d\om \,\om \sqrt{\om^2-m_1^2} \,
G^{T}_{\chi} (\om ;\tau,\tau)
+\frac{\l N_2}{2\pi^2}\int_{m_2}d\om \,\om \sqrt{\om^2-m_2^2} \,
G^{T}_{\phi} (\om ;\tau,\tau) .~~~~~
\label{uuvsG}
\eea
The derivation uses factorization at large $N$.

\subsection{Unbroken reflection symmetry}

The solution to Eqs.~\rf{eeq1}--\rf{uuvsG} is very similar to that of
Subsect.~2.3, while there are important differences. It reads
\be
 G^{T}_{\chi} (\om ;\tau,\tau)
= \frac{1}{2\om} - \frac{\bar{\l}_1\Phi^2}{4\om (\om^2 -m_{1R}^2)}
-\frac{\bar{\l}_2\Psi^2}{4\om (\om^2 -m_{2R}^2)}\,,
\label{dres}
\ee
where
\be
v_R = \bar{\l}_1 \Phi^2 + \bar{\l}_2 \Psi^2 \,,
\ee
\be
\bar{\l}_1 = \frac{\l_R}{1+\frac{\l_R}{8\pi^2}I_1}~,~~~~~
\bar{\l}_2 = \frac{\l_R}{1+\frac{\l_R}{8\pi^2}I_2}~,
\label{llbar}
\ee
and
\be
I_1= N_1 + N_2 \int_{m_{2R}}
d\om \sqrt{\om^2 -m_{2R}^2}
\left[ \frac{1}{\om^2 - m_{1R}^2} -\frac{1}{\om^2} \right],
\label{I1}
\ee
\be
I_2= N_2 + N_1 \int_{m_{1R}}
d\om \sqrt{\om^2 -m_{1R}^2} \left[ \frac{1}{\om^2 - m_{2R}^2}
-\frac{1}{\om^2} \right] ,
\label{I2}
\ee
while $\Phi^a$ and $\Psi^\a$ are similar to those of Ref.~\cite{LRT93a}:
\bea
{\Phi}^a =
{z}_1^a \left( 1-2\bar{\l}_2 \frac{\ka}{m_{2R}^2} {z}^2_2 \right)
\left( 1- \frac{2\bar{\l}_1}{m_{1R}^2} {z}^2_1
-\frac{2\bar{\l}_2}{m_{2R}^2} {z}^2_2
+ \bar{\l}_1\bar{\l}_2 \frac{\ka^2}{m^2_{1R} m^2_{2R}} {z}_1^2 {z}_2^2
\right)^{-1} , \\
{\Psi}^{\b} =
{z}_2^{\b} \left( 1+2\bar{\l}_1 \frac{\ka}{m_{1R}^2} {z}^2_1 \right)
\left( 1- \frac{2\bar{\l}_1}{m_{1R}^2} {z}^2_1
-\frac{2\bar{\l}_2}{m_{2R}^2} {z}^2_2
+ \bar{\l}_1\bar{\l}_2 \frac{\ka^2}{m^2_{1R} m^2_{2R}} {z}_1^2 {z}_2^2
\right)^{-1} ,
\label{rubF}
\eea
with
\be
\ka=\frac{m_{1R}-m_{2R}}{m_{1R}+m_{2R}}
\ee
and
\be
{z}_1^a = \xi_1^a  e^{m_{1R} \tau}~,~~~~~
{z}_2^{\a} = \xi_2^{\a}  e^{m_{2R} \tau} \,.
\ee

\begin{figure}[tbp]
\unitlength=1.00mm
\linethickness{0.6pt}
\centering
\begin{picture}(118.00,38.00)(11,90)
\put(70.00,114.00){\circle{14.00}}
\multiput(77.50,114.00)(7.0,3.50){3}{
    \line(2,1){4.00}}
\multiput(77.50,114.00)(7.0,-3.50){3}{
    \line(2,-1){4.00}}
\put(61.00,114.00){\line(-2,-1){18.00}}
\put(61.00,114.00){\line(-2,1){18.00}}
\end{picture}
\caption[x]   {\hspace{0.2cm}\parbox[t]{13.5cm}
{\small
   The one-loop diagram for the process $\chi+\chi \ra \phi+\phi$
   which possesses an imaginary part for $m_{2R}>m_{1R}$
   given by \eq{IMG}. }}
\label{fig55}
\end{figure}
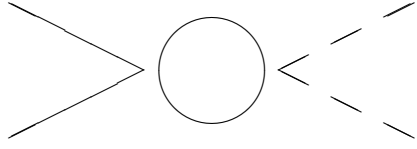

The constants $\bar{\l}_1$ and $\bar{\l}_2$ are complex in general.
Assuming for definiteness that $m_{2R}^2 \geq m_{1R}^2$, one can easily see
that the constant $\bar{\l}_1$ is real while the imaginary part of
$\bar{\l}_2$ is given by
\be
{\rm Im}\, \frac{1}{\bar{\l}_2} = -\frac{N_1}{8\pi} \, \frac{\sqrt{m_{2R}^2
-m_{1R}^2}}{2m_{2R}} .
\label{IMG}
\ee
The appearance of the imaginary part
is because the masses of the $O(N_1)$ and $O(N_2)$ particles are not equal.
For $m_{2R}^2 \geq m_{1R}^2$ the imaginary part is associated
with an inelastic process $\chi +\chi \ra \phi +\phi$ which is given
to lowest order by the diagram depicted Fig.~\ref{fig55}.

\subsection{Broken reflection symmetry}

For $N_2=1$ when $\phi$ is a singlet field, the symmetry of the
Lagrangian~\rf{llagrangian} under the reflection $\phi\ra-\phi$ is
spontaneously broken if $m_2^2<0$. The remaining $O(N_1)=O(N$$-$$1)$
symmetry is unbroken for $m_1^2\geq-|m_2^2|$ (for $m_1^2=m_2^2<0$ the whole
$O(N)$ symmetry were spontaneously broken down to $O(N$$-$$1)$ and
the associated massless Goldstone bosons would appear).

The vacuum expectation value of the field $\phi$ and
the physical mass of the field $\chi$ are
\be
\eta_R = \frac{|m_{2R}|}{\sqrt{\l_R}}~,~~~~~~m_\chi^2 = m_{1R}^2-m_{2R}^2=
m_1^2-m_2^2\,,
\label{masses}
\ee
while the physical mass of the shifted field $\phi^\p=\phi-\eta_R$
is determined self-consistently by
\be
m_\phi^2 = 2\bar{\l}_2 \eta^2_R
\ee
with $\bar{\l}_1$, $\bar{\l}_2$ given by \eq{llbar} and
\bea
I_1& =& N \int_{m_{\chi}} d\om
\sqrt{\om^2 -m_{\chi}^2} \left[ \frac{\om^2}{(\om^2 -
m_\phi^2/4)(\om^2 -m_{\chi}^2)}
- \frac{1}{\om^2} \right] , \\
I_2 &=& N \int_{m_{\chi}} d\om
\sqrt{\om^2 -m_{\chi}^2} \left[ \frac{1}{\om^2 -m_\phi^2/4}
- \frac{1}{\om^2} \right] .
\eea
The parameters $\bar{\l}_1$ and $\bar{\l}_2$ are real
in contrast
to the case without the spontaneous breaking of the
reflection symmetry.

The solution reads
\be
 G^{T}_\chi(\om;\tau,\tau) = \frac{\om}{2(\om^2 -m_\phi^2/4)} -
\frac{\bar{\l}_1 \om \Phi^2}{4(\om^2 - m_{\phi}^2/4) (\om^2-m_{\chi}^2)}
-\frac{\bar{\l}_2 \Psi^2}{4\om(\om^2 - m_{\phi}^2/4)},
\label{resB}
\ee
\be
v_R=\bar{\l}_1 \Phi^2 +\bar{\l}_2 \Psi^2 -\frac{m_\phi^2}{2}\,,
\label{vB}
\ee
where
\be
\Phi^a  = z_1^a \,\frac{1-\frac{(2m_{\chi} -m_{\phi})}{(2m_{\chi}
+m_{\phi})} \frac{{z}_2}{2\eta_{R}}}
{1 - \frac{z_2}{2\eta_{R}} -
\frac{8\bar{\l}_1}{4m_{\chi}^2 - m_{\phi}^2} {z}_1^2 +
\frac{4\bar{\l}_1}{\eta_{R}} \frac{(2m_{\chi} -m_{\phi})}{(2m_{\chi} +
m_{\phi})^3} {z}_2 {z}^2_1 } \,,
\ee
\be
\Psi = \eta_R+ \eta_R \, \frac{ \frac{z_2}{\eta_R} +
\frac{16\bar{\l}_1}{4m_{\chi}^2 - m_{\phi}^2} {z}_1^2 }
{ 1 - \frac{{z_2}}{2\eta_{R}} -
\frac{8\bar{\l}_1}{4m_{\chi}^2 - m_{\phi}^2} {z}_1^2 +
\frac{4\bar{\l}_1}{\eta_{R}}
\frac{(2m_{\chi} -m_{\phi})}{(2m_{\chi} + m_{\phi})^3} {z}_2 {z}^2_1}
\ee
and
\be
{z}_1^a =m_{\chi} \xi_1^a e^{m_{\chi} \tau}~,~~~~
{z}_2 = m_{\phi} \xi_2 e^{m_{\phi}\tau} \,.
\ee
This solution is similar to the tree level one~\cite{LRT93a} and is real.

The poles of~\rf{resB} at $\om =m_{\chi}$ and $\om =m_{\phi}/2$
are associated with production of two
$\chi$-particles or a single $\phi$ particle, respectively.
There is no pole at $\om =m_{\phi}$
which signals that the amplitude for the production of two
$\phi$ particles vanishes in contrast to the cases without
the spontaneous breaking and $N=1$ with broken reflection
symmetry~\cite{Smi93}.
It is easy to see shifting $\Psi=\eta_R+\Psi^\p$ that~\rf{resB} and~\rf{vB}
recover the free values $1/2\om$ and $0$, respectively, to zeroth order of
perturbation theory in $\l_R$ around the shifted vacuum.

\newsection{Generalizations}

\subsection{Gelfand--Diki\u{\i} technique for P\"{o}schl--Teller potential
\label{gd}}

The results presented in Sects.~2, 3 are obtained by the
Gelfand--Diki\u{\i} technique~\cite{GD75} which represents the
{\it diagonal\/} resolvent of the Schr\"{o}dinger operator in \eq{eqforG} as
\be
G^{T}_\om(\tau,\tau)=
 R_\om[v] \equiv \sum_{l=0}^\infty
\frac{R_l[v]}{\om^{2l+1}}
\label{GD}
\ee
where the differential polynomials $R_l[v]$ are determined
recurrently by
\be
R_{l+1}[v] = \fr{1}{2} \left( \fr 12 D^2-v- D^{-1}vD\right) R_l[v]~,~~~~~
R_0=\fr 12
\label{polynomial}
\ee
and the inverse operator is
\be
D^{-1} f(\tau) = \int^\tau_{-\infty} dx \;f(x) \,.
\ee
\eq{polynomial} stems from the fact that $R_\om[v]$ obeys
the third  order linear differential equation
\be
\fr 12 \left( \fr 12 D^3 -Dv -vD \right)R_\om[v] =
\om^2 DR_\om[v]\,.
\label{linear}
\ee

The potential~\rf{exact1}, \rf{finalPhi} is a particular case of a more general
P\"{o}schl--Teller potential
\be
v(\tau)=m^2\left\{\frac{s(s+1)}{\sinh^2{(m(\tau-\tau_0))}}
- \frac{k(k+1)}{\cosh^2{(m(\tau-\tau_0))}}
\right\}
\label{PT}
\ee
with $s=1$ and $k=0$.
The simple answer~\rf{exact} for the diagonal resolvent
is because
\be
R_l[v]= m^{2l-2}R_1[v]=-\frac 14 m^{2l-2} v~~~~~~~~\hbox{for}~~~~~s=1,k=0 \,,
\label{reduce}
\ee
\ie $R_l[v]$ with $l>1$ reduces to $R_1[v]$.

It is easy to extend Eqs.~\rf{exact}, \rf{exact1}
to case $s>1$, $k=0$ (or $k>1$, $s=0$)
when
\be
R_\om[v] = \sum_{l=0}^s
\frac{C_l\, v^l}{\om(\om^2- m^2)\cdots(\om^2-l^2 m^2)}
\label{ansatz}
\ee
and the coefficients $C_l$ are determined recurrently by
\be
C_{l+1}=\frac{l(l+1)-s(s+1)}{(l+1)s(s+1)} \left( l+\fr 12\right) C_l \,.
\label{recurrenceB}
\ee
These formulas can easily be proven substituting into \eq{linear} and
noticing that
\be
\fr{1}{2} \left( \fr 12 D^2-v-
D^{-1}vD \right) v^l
 = l^2 m^2 v^l +
\frac{l(l+1)-s(s+1)}{(l+1)s(s+1)} \left( l+\fr 12\right) v^{l+1}\,.
\label{eigenvector}
\ee
It is evident from \eq{eigenvector} that $v^s$, which is the highest power of
$v$ in~\rf{ansatz}, is an eigenvector of the operator entering \eq{polynomial}
and, therefore, $R_l[v]$ for $l>s$ reduces to $R_1[v],\ldots,R_s[v]$.

The simple form~\rf{ansatz} of the diagonal resolvent for the potential~\rf{PT}
with $k=0$ is due to the fact that the spectrum is very simple:
there are $s$ poles at $\om=1,\ldots,s$ (in the units of $m$).
For nonvanishing both $s$ and $k$%
\footnote{~The potential~\rf{PT} with $k=1$ and arbitrary $s$
(which is related to the ratio of fermion to vector-boson masses)
appears in the standard model
for longitudinally polarized vector bosons~\cite{AKP93d,Smi94}.}
the spectrum is slightly more complicated:
there are $M=\hbox{max}\,\{s,k\}$ poles which go from $1$ up to $|s-k|$
with the step $1$ and then from $|s-k|+2$ up to $s+k$ with the step $2$.
Say, the spectrum for $s=2$ and $k=5$ is $1,2,3,5,7$.
This spectrum can be
obtained from the one for $k=0$ (or $s=0$) by a supersymmetry transformation%
\footnote{~I am grateful to E.~Argyres and A.~Johansen for discussions of
this point.}
which relates $(s,k)$ with $(s\!+\!1,k\!+\!1)$.
It was explicitly done for $k=1$ and any $s$ by Smith~\cite{Smi94}.

For the potential~\rf{PT} $R_l[v]$ with
$l>M=\hbox{max}\,\{s,k\}$ reduces to $R_1[v],\ldots,R_M[v]$.
This is based on the fact that
\be
\fr 12 \left( \fr 12 D^2 -v -D^{-1}vD \right)
\sinh^{-2s}{\tau}\,\cosh^{-2k}{\tau} =
(s+k)^2 \sinh^{-2s}{\tau}\,\cosh^{-2k}{\tau}
\label{highest}
\ee
for the potential~\rf{PT} (I put $m=1$ and $\tau_0=0$ for brevity),
\ie there exists an eigenvector of the operator entering
the recurrence relation~\rf{polynomial}.
As is shown in Ref.~\cite{Mak94}, if $\Phi(\tau)$ is an eigenvector of the
Schr\"{o}dinger operator on the l.h.s.\ of \eq{eq1} with an eigenvalue $E$
and $\Phi(-\infty)=0$, then $\Phi^2(\tau)$ will be an eigenvector of the
operator on the l.h.s.\ of \eq{highest} with the same eigenvalue. Therefore,
spectra of the two operators coincide.

Since the spectrum is known, a modification of~\rf{ansatz} for $k\neq0$ is
obvious while awkward.  An analog of \eq{reduce} can be obtained as follows.
For the given spectrum $m_1,\ldots,m_M$ let us calculate
\be
{\cal D}^{(s,k)}(\om) \equiv \prod_{i=1}^{M} (\om^2 - m_i^2) =
\om^{2M} - \sum_{i=1}^M D_i^{(s,k)} \om^{2(M-i)}
\ee
to determine the coefficients $D^{(s,k)}_i$.
Then, the reduction of $R_{M+1}[v]$ reads
\be
R_{M+1}=\sum_{i=1}^M D^{(s,k)}_i R_{M-i+1} \,.
\label{reduceM}
\ee
The reduction of $R_l[v]$ for $l>M$$+$$1$ can be obtained from \eq{reduceM}
recurrently using \eq{polynomial}.

Let us illustrate \eq{reduceM} which relies on \eq{highest} by an example
of $(s$$=$$2, k$$=$$5)$ with the spectrum mentioned above. One gets
\bea
{\cal D}^{(2,5)}(\om)=(\om^2-1) (\om^2-4)(\om^2-9)(\om^2-25)(\om^2-49)= \non
\om^{10}-88 \om^8 + 2310 \om^6 -20812\om^4+ 62689 \om^2 - 44100
\eea
and
\be
R_6=88 R_5 - 2310 R_4 +20812 R_3- 62689 R_2 + 44100 R_1 \,,
\ee
while the diagonal resolvent is
\bea
&R_\om[v]=\frac{1}{2\om} + \frac{R_1}{\om(\om^2-1)}
+ \frac{R_2-R_1}{\om(\om^2-1)(\om^2-4)}
+ \frac{R_3-5R_2+4R_1}{\om(\om^2-1)(\om^2-4)(\om^2-9)}&\non&
+ \frac{R_4-14R_3+49R_2-36R_1}{\om(\om^2-1)(\om^2-4)(\om^2-9)(\om^2-25)}
+ \frac{R_5-39R_4+399R_3-1261R_2+900R_1}{\om(\om^2-1)(\om^2-4)(\om^2-9)
(\om^2-25)(\om^2-49)} \,.&
\label{25}
\eea
The coefficients of the numerators in the last term on the r.h.s.\
are just $D_i^{(1,4)}$ associated with the spectrum $1,2,3,5$ for
$(s$$=$$1,k$$=$$4)$ as is prescribed by the supersymmetry.
Analogously, the numerators of the 5th, 4th and 3rd terms on the r.h.s.\
correspond to the $(s$$=$$0,k$$=$$3)$, $(s$$=$$0,k$$=$$2)$
and $(s$$=$$0,k$$=$$1)$ spectra, respectively.
Substituting~\rf{PT} and using {\it Mathematica\/}, \rf{25} can be
written explicitly as
\bea
&R_\om[v]=\frac{1}{2\om} +
\frac{3\sinh^{-2}{\tau}(4+\cosh^{-2}{\tau})} {2\om(\om^2-1)} +
\frac{3^2\sinh^{-4}{\tau}(21-55\cosh^{-2}{\tau}+35\cosh^{-4}{\tau})}
{2\om(\om^2-1)(\om^2-4)}&\non&
+ \frac{3^2\cdot5^2\sinh^{-4}{\tau}\cosh^{-2}{\tau}
(10-35\cosh^{-2}{\tau}+28\cosh^{-4}{\tau})}{2\om(\om^2-1)(\om^2-4)(\om^2-9)}
& \non &+
\frac{3^3\cdot5^2\cdot7^2\sinh^{-4}{\tau}\cosh^{-8}{\tau}}
{2\om(\om^2-1)(\om^2-4)(\om^2-9)(\om^2-25)}
+ \frac{3^6\cdot5^2\cdot7^2\sinh^{-4}{\tau}\cosh^{-10}{\tau}}
{2\om(\om^2-1)(\om^2-4)(\om^2-9)
(\om^2-25)(\om^2-49)} \,.&
\eea
The highest vector is in agreement with \eq{highest}.

\subsection{Symmetrized amplitude and renormalons}

\begin{figure}[tbp]
\unitlength=1.00mm
\linethickness{0.6pt}
\centering
%\begin{picture}(118.00,68.00)(20,70)
\begin{picture}(118.00,68.00)(15,70)
\put(23.00,78.00){\makebox(0,0)[cc]{{\Large a)}}}
\put(5.00,128.00){\line(1,0){18.00}}
\put(5.00,100.00){\line(1,0){18.00}}
\put(23.00,101.50){\line(0,1){25.00}}
\put(23.00,128.00){\line(2,1){12.00}}
\put(23.00,100.00){\line(2,-1){12.00}}
\put(23.00,126.50){\line(2,-1){12.00}}
\put(23.00,101.50){\line(2,1){12.00}}
%2
\put(70.00,78.00){\makebox(0,0)[cc]{{\Large b)}}}
\put(70.00,114.00){\line(-2,-1){18.00}}
\put(70.00,115.50){\line(-2,1){18.00}}
\put(70.00,115.50){\line(2,1){18.00}}
\put(70.00,114.00){\line(1,-1){18.00}}
\put(80.00,105.50){\line(4,-1){8.00}}
\put(80.00,105.50){\line(2,1){8.00}}
%3
\put(125.00,119.50){\line(-2,1){18.00}}
\put(125.00,119.50){\line(2,1){18.00}}
\put(125.00,108.50){\line(-2,-1){18.00}}
\put(125.00,108.50){\line(2,-1){18.00}}
\put(125.00,78.00){\makebox(0,0)[cc]{{\Large c)}}}
\put(125.00,114.00){\circle{8.00}}
\put(130.50,114.00){\line(2,1){12.50}}
\put(130.50,114.00){\line(2,-1){12.50}}
\end{picture}
\caption[x]   {\hspace{0.2cm}\parbox[t]{13.5cm}
{\small
   Some large-$N$ diagrams for the symmetrized amplitude $G^{S}(4;p)$. }}
   \label{fig6}
   \end{figure}
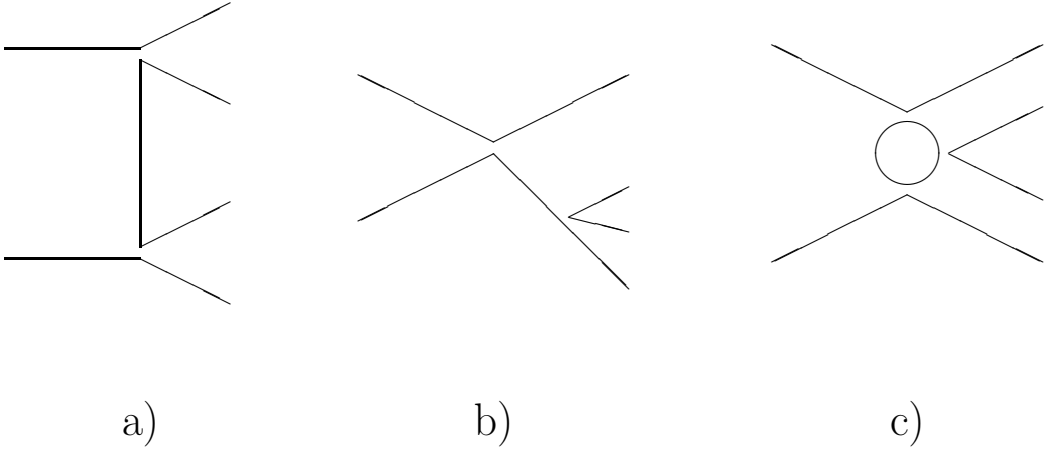

The symmetrized amplitude
\be
D^S(n;p)  = D^{ab}(n;p)\frac{\xi^a \xi^b}{\xi^2}
\label{GS}
\ee
reveals, on the contrary, a
surprisingly nontrivial behavior even at large $N$.
Some of the diagrams which contribute to $D^S(4;p)$
are depicted in Fig.~\ref{fig6}.
The combinatorics is now
different from that of the diagrams of Fig.~\ref{fig5}
so  there is no cancellation of tree diagrams
for $D^S(4;p)$ on mass shell which happens for $n$$\geq$$6$
according to the explicit result~\cite{Smi93a} for $D^S(4;p)$
at the tree level.
Each of the vertices can be
``dressed'' with bubble chains and each of the lines
of outgoing particles can be substituted by the exact
amplitude $a^b(n)$ to obtain a diagram with more
produced particles.

\begin{figure}[tbp]
\unitlength=1.00mm
\linethickness{0.6pt}
\centering
%\begin{picture}(118.00,68.00)(20,70)
\begin{picture}(118.00,72.00)(15,70)
\put(23.00,78.00){\makebox(0,0)[cc]{{\Large a)}}}
\put(23.00,128.00){\line(-5,-4){17.50}}
\put(23.00,100.00){\line(-5,4){17.50}}
\put(23.00,101.50){\line(0,1){25.00}}
\put(23.00,128.00){\line(2,1){12.00}}
\put(23.00,100.00){\line(2,-1){12.00}}
\put(23.00,126.50){\line(2,-1){12.00}}
\put(23.00,101.50){\line(2,1){12.00}}
%2
\put(75.00,78.00){\makebox(0,0)[cc]{{\Large b)}}}
\put(75.00,132.00){\line(-5,-4){22.50}}
\put(75.00,96.00){\line(-5,4){22.50}}
\put(77.00,111.50){\line(0,1){5.00}}
\put(75.00,132.00){\line(2,1){12.00}}
\put(75.00,96.00){\line(2,-1){12.00}}
\put(77.00,111.50){\line(1,0){10.00}}
\put(77.00,116.50){\line(1,0){10.00}}
\put(77.00,129.00){\circle{5.00}}
\put(77.00,120.00){\circle{5.00}}
\put(77.00,99.00){\circle{5.00}}
\put(77.00,108.00){\circle{5.00}}
\put(77.00,104.60){\makebox(0,0)[cc]{$\vdots$}}
\put(77.00,125.50){\makebox(0,0)[cc]{$\vdots$}}
%3
\put(125.00,132.00){\line(-5,-4){22.50}}
\put(125.00,96.00){\line(-5,4){22.50}}
\put(127.00,114.00){\circle{5.00}}
\put(125.00,132.00){\line(2,1){12.00}}
\put(125.00,96.00){\line(2,-1){12.00}}
\put(130.50,114.00){\line(2,1){6.50}}
\put(130.50,114.00){\line(2,-1){6.50}}
\put(127.00,129.00){\circle{5.00}}
\put(127.00,120.00){\circle{5.00}}
\put(127.00,99.00){\circle{5.00}}
\put(127.00,108.00){\circle{5.00}}
\put(127.00,104.60){\makebox(0,0)[cc]{$\vdots$}}
\put(127.00,125.50){\makebox(0,0)[cc]{$\vdots$}}
\put(125.00,78.00){\makebox(0,0)[cc]{{\Large c)}}}
\end{picture}
\caption[x]   {\hspace{0.2cm}\parbox[t]{13.5cm}
{\small
   The diagrammatic representation of some diagrams for $D^{S}(4;p)$
   at large $N$ integrated over $d^4 p$.
   The diagrams b) and c) are of the renormalon type.
   }}
   \label{fig7}
   \end{figure}
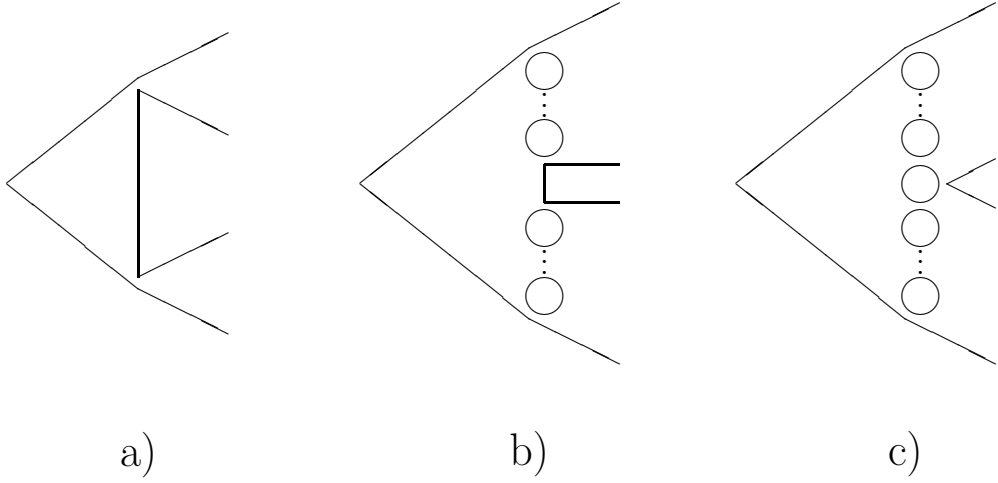

One can integrate $D^S(4;p)$ over the $4$-momentum $p$
to obtain diagrams of the vertex type. Some of them are
depicted in Fig.~\ref{fig7}.
The most interesting are the diagrams
of Fig.~\ref{fig7}b),~c) which are of the type of
renomalons and should behave to $k$-th
order of perturbation theory as $k!$.

In order to calculate $D^{S}(n;p)$ for the unbroken $O(N)$
symmetry at large $N$,
we first derive the proper Schwinger--Dyson equation substituting
$F=\phi^b(\vec{0},-i\tau^\p)$ in \eq{identity} and symmetrizing according to
\eq{GS}. The resulting equation can be written as
\bea
& &\left\{\frac{d^2}{d\tau^2} -\om^2- v(\tau) - 2\l \Phi^2(\tau)\right\}
G^{S}_\om(\tau,\tau^\p) \non & & - 2 \l\Phi^a(\tau)
\int \frac{d^3 \vec{p}}{(2\pi)^3}\e^{i(\vec{p}_1+\vec{p}_2) \vec{x}}
\LA \Xi| \phi^2(\vec{x},-i\tau) \phi^a(\vec{0},-i\tau)|0 \RA_{conn}
%\int \frac{d \epsilon}{2\pi} \e^{-\epsilon \tau^\p}
%\int \frac{d^4 k}{(2\pi)^4} D^a(\tau;p,p-k)
= - \delta(\tau-\tau^\p) \,. ~~~~~~~~
\label{eqforGS}
\eea
This equation involves a new amplitude
\be
D^a(\tau;p_1,p_2) =\frac{1}{N} \int \frac{d^4p_1}{(2\pi)^4}
\frac{d^4p_2}{(2\pi)^4} \e^{ip_1x+ip_2y-(\epsilon_1+\epsilon_2)\tau}
\LA \Xi| \phi^b(x)\phi^b(y) \phi^a(\vec{0},-i\tau)|0 \RA_{conn}
\ee
with three external lines having nonvanishing spatial momenta
(which is a generating function for $D^a(n;p_1,p_2)$),
where $a$ is the $O(N)$ index of an outgoing particle with the spatial
momentum $\vec{p}_1+\vec{p}_2$ while the averaging over the $O(N)$ indices
of two incoming particles with the spatial momenta $\vec{p}_1$ and $\vec{p}_2$
is performed.

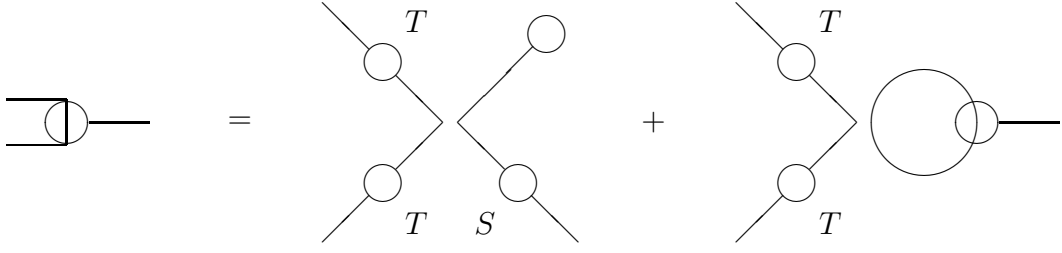
\begin{figure}[tp]
\unitlength=1.00mm
\linethickness{0.6pt}
\centering
\begin{picture}(133.00,45.00)(0,95)
\put(4.00,120.00){\circle{6.00}}
\put(7.00,120.00){\line(1,0){8.00}}
\put(4.00,123.00){\line(-1,0){8.00}}
\put(-4.00,117.00){\line(1,0){8.00}}
\put(4.00,117.00){\line(0,1){6.00}}
\put(27.00,120.00){\makebox(0,0)[cc]{$=$}}
\put(54.00,120.00){\line(-1,1){6.22}}
\put(54.00,120.00){\line(-1,-1){6.22}}
\put(46.00,128.00){\circle{5.00}}
\put(46.00,112.00){\circle{5.00}}
\put(44.22,129.78){\line(-1,1){6.22}}
\put(44.22,110.22){\line(-1,-1){6.22}}
\put(49.00,132.00){\makebox(0,0)[lb]{$T$}}
\put(49.00,108.00){\makebox(0,0)[lt]{$T$}}
\put(56.00,120.00){\line(1,-1){6.22}}
\put(64.00,112.00){\circle{5.00}}
\put(65.78,110.22){\line(1,-1){6.22}}
\put(61.00,108.00){\makebox(0,0)[rt]{$S$}}
\put(56.00,120.00){\line(1,1){10.00}}
\put(67.78,131.78){\circle{5.00}}
\put(82.00,120.00){\makebox(0,0)[cc]{$+$}}
\put(109.00,120.00){\line(-1,1){6.22}}
\put(109.00,120.00){\line(-1,-1){6.22}}
\put(101.00,128.00){\circle{5.00}}
\put(101.00,112.00){\circle{5.00}}
\put(99.22,129.78){\line(-1,1){6.22}}
\put(99.22,110.22){\line(-1,-1){6.22}}
\put(104.00,132.00){\makebox(0,0)[lb]{$T$}}
\put(104.00,108.00){\makebox(0,0)[lt]{$T$}}
\put(118.00,120.00){\circle{14.00}}
\put(125.00,120.00){\circle{6.00}}
\put(128.00,120.00){\line(1,0){8.00}}
\end{picture}
\caption[x]   {\hspace{0.2cm}\parbox[t]{13.5cm}
{\small
   The graphic representation of \eq{sdva}.
   The vertex on the l.h.s.\
   represents $D^a(\tau;p_1,p_2)$, the circles with two lines represent
   $D^T(\tau;p)$ or $D^S(\tau;p)$ and the tadpole represents $\Phi^a(\tau)$.
    }}
\label{sdv}
\end{figure}

The Scwhinger--Dyson equation  which is satisfied by $D^a(\tau;p_1,p_2)$
is now of the vertex type rather than of the propagator type as before.
It is depicted graphically in Fig.~\ref{sdv}%
\footnote{ This equation differs at the tree level from the one of
Ref.~\cite{Smi93b} since we use exact propagators which are ``dressed'' by
emitting on-mass-shell particles.}
and reads analytically as
\bea
D^a(\tau;p_1,p_2) & =& \l N D^T(\tau;p_1) D^T(\tau;p_2)
\non & & \times\left[
\Phi^a D^S(\tau;p_1+p_2) + \int \frac{d^4 k}{(2\pi)^4}
D^a(\tau;k,p_1+p_2-k) \right].
\label{sdva}
\eea
Its solution for $D^a(\tau;p_1,p_2)$ is given by
\be
D^a(\tau;p_1,p_2) = \frac{
\l N D^T(\tau;p_1) D^T(\tau;p_2) \Phi^a(\tau) D^S(\tau;p_1+p_2)}
{1-\l N\int \frac{d^4 k}{(2\pi)^4} D^T(\tau;k)
D^T(\tau;p_1+p_2-k)}\,.
\label{loopintegral}
\ee

The propagator $D^T(\tau;p)$ is determined by \eq{eqforG} and reads
explicitly
\bea
&D^T(\tau;p)=\int \frac{d\tau}{i} \e^{\epsilon(\tau^\p-\tau)}
G_\om(\tau,\tau^\p)  =
\frac{i}{\epsilon^2-\om^2}\left[ 1+\frac{2m(m+\epsilon)}{\om^2-m^2}\right]
&\non& - \frac{2i}{\om^2-m^2}
\sum_{n=0}^\infty \left( \frac{\bar{\l}\xi^2}{8}\e^{-2m\tau} \right)^n
\frac{\epsilon m-2nm^2}{(\epsilon-2nm)^2-\om^2}&\non&
+\frac{2i}{\om^2-m^2}\left(1-\frac{2}{1-\frac{\bar{\l}\xi^2}{8}\e^{-2m\tau}}
\right)
\sum_{n=0}^\infty \left( \frac{\bar{\l}\xi^2}{8}\e^{-2m\tau} \right)^n
\frac{m^2}{(\epsilon-2nm)^2-\om^2}&
\label{Genergy}
\eea
where the integral over $d\tau$ is understood as an analytic continuation
from imaginary $\tau$ (cf.\ \rf{Dphi}).
The expression~\rf{Genergy} can be simplified using the formula
\be
\sum_{n=0}^\infty \frac{z^n}{n+u} = u^{-1} \,{}_2 \hbox{F}_1(1,u;1+u;z)\,.
\ee

The calculation of the renormalon contribution would consist of three steps.
\begin{itemize} \vspace{-6pt}
\addtolength{\itemsep}{-7pt}
\item[1.]
Calculate the integral $\int
\frac{d^4 k}{(2\pi)^4} D^T(\tau;k) D^T(\tau;p-k)$.

\item[2.]
Substutute the (proper Fourier transformed) expression~\rf{loopintegral}
into \eq{eqforGS} and solve it for the diagonal
resolvent $G^S_\om(\tau,\tau)$.

\item[3.]
Integrate $\int_{m^2} d\om\, \om\,\sqrt{\om^2-m^2}\; G^S_\om(\tau,\tau)$.
\vspace{-4pt}
\end{itemize}

\subsection{Matrix Higgs field}

Another interesting model to investigate is the
case of the {\it matrix\/} Higgs field which is described
by the Lagrangian
\be
{\cal L} = \frac 12 \tr{ (\partial_\mu \phi) ^2}
-\frac{m^2}{2}  \tr {\phi ^2}
-\frac{\l_3}{3} \tr {\phi ^3}-
\frac{\l_4}{4}  {\tr \phi ^4}\,,
\label{mlagrangian}
\ee
where $\phi^{ij}(x)$ is generically $N$$\times$$N$ Hermitean matrix.
The model greatly simplifies as $N\ra \infty$ at fixed
$\l_3^2 {N}$ or $\l_4 N$ when only the planar diagrams
survive  similar to  the 't~Hooft large-$N$ limit of QCD.

\begin{figure}[tb]
\unitlength=1.00mm
\linethickness{0.6pt}
\centering
%\begin{picture}(153.00,68.00)(5,70)
\begin{picture}(151.00,68.00)(6,70)
\put(23.00,78.00){\makebox(0,0)[cc]{{\Large a)}}}
\put(7.00,128.00){\line(1,0){15.00}}
\put(7.00,100.00){\line(1,0){15.00}}
\put(22.00,100.00){\line(0,1){28.00}}
\put(7.00,130.00){\line(1,0){32.00}}
\put(7.00,98.00){\line(1,0){32.00}}
\put(24.00,128.00){\line(1,0){15.00}}
\put(24.00,100.00){\line(1,0){15.00}}
\put(24.00,115.00){\line(1,0){15.00}}
\put(24.00,113.00){\line(1,0){15.00}}
\put(24.00,128.00){\line(0,-1){13.00}}
\put(24.00,100.00){\line(0,1){13.00}}
\put(72.00,78.00){\makebox(0,0)[cc]{{\Large b)}}}
\put(53.00,128.00){\line(1,0){15.00}}
\put(53.00,100.00){\line(1,0){15.00}}
\put(68.00,100.00){\line(0,1){28.00}}
\put(53.00,130.00){\line(1,0){27.00}}
\put(53.00,98.00){\line(1,0){32.00}}
\put(70.00,128.00){\line(1,0){10.00}}
\put(70.00,100.00){\line(1,0){15.00}}
\put(70.00,128.00){\line(0,-1){28.00}}
\put(80.00,130.00){\line(2,1){12.00}}
\put(80.00,128.00){\line(2,-1){12.00}}
\put(82.40,129.00){\line(2,-1){9.60}}
\put(82.40,129.00){\line(2,1){9.60}}
\put(128.00,78.00){\makebox(0,0)[cc]{{\Large c)}}}
\put(120.60,114.00){\line(-2,-1){15.00}}
\put(120.60,114.00){\line(-2,1){15.00}}
\put(123.00,115.00){\line(-2,1){17.40}}
\put(123.00,113.00){\line(-2,-1){17.40}}
\put(123.00,115.00){\line(1,0){10.00}}
\put(123.00,113.00){\line(1,0){10.00}}
\put(133.00,115.00){\line(2,1){17.40}}
\put(135.40,114.00){\line(2,1){15.00}}
\put(133.00,113.00){\line(2,-1){8.60}}
\put(133.00,113.00){\line(2,-1){22.60}}
\put(135.40,114.00){\line(2,-1){9.60}}
\put(145.00,109.20){\line(2,1){10.60}}
\put(147.20,108.10){\line(2,-1){8.40}}
\put(147.20,108.10){\line(2,1){8.40}}
\end{picture}
\caption[x]   {\hspace{0.2cm}\parbox[t]{13.5cm}
{\small
   The tree diagrams for a $2$$\ra$$3$ amplitude
   in matrix $\phi^3$ theory at large $N$. The continuous lines
   are associated with propagation of matrix indices.
   The contributions of the diagrams a), b) and c) equal, respectively,
   $1/4$, $-1/6$ and $1/24$ for the given on-mass-shell kinematics
   when the momenta of the incoming particles (in the units of mass)
   are $(3/2,\sqrt{5}/2,0,0)$ and $(3/2,-\sqrt{5}/2,0,0)$,
   while that of the each produced particle is $(1,0,0,0)$.
   The diagrams b) and c) have an extra combinatorial factor $2$ each.
   The sum of three diagrams vanishes which illustrates the nullification
   of $2$$\ra$$3$ on mass shell at the tree level.
   }}
\label{phi3}
\end{figure}
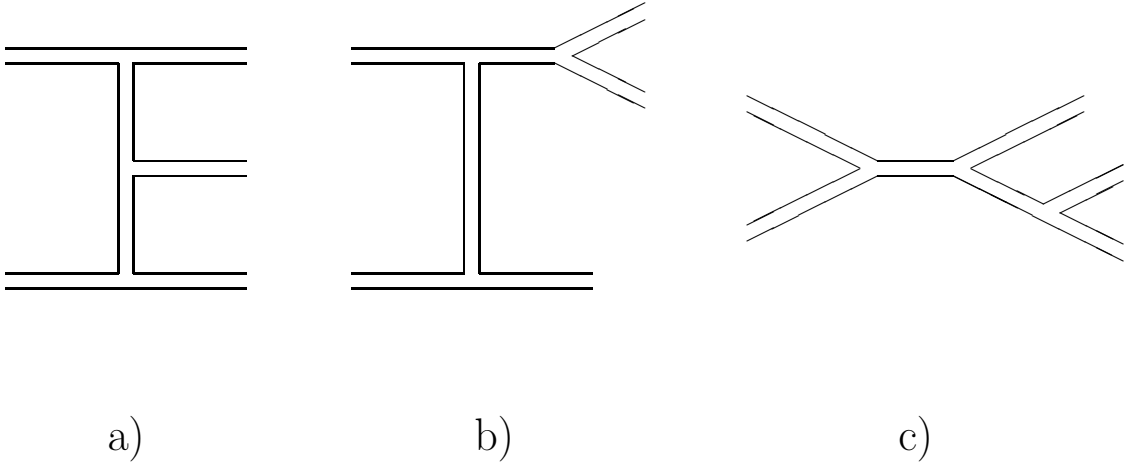

The nullification of on-mass-shell
amplitudes holds in the case of the matrix cubic interaction
for $2$$\ra$$3$ at large $N$ at the tree level.
The cancellation of diagrams is illustrated by Fig.~\ref{phi3}.
This looks quite similar to the $O(N)$-vector case.

The large-$N$ limit of the matrix $\phi^4$ theory
while being simpler than the $N=1$ case to all loops
is, however,
quite nontrivial like the large-$N$ limit of QCD.
Some of the rescattering diagrams are now included in the large-$N$
amplitudes so they should not look like the tree-level ones.
The crucial problem is to find a simple set of observables like $\tr{\phi^4}$
for ordinary matrix models to close the Schwinger--Dyson equations for the
given kinematics at large $N$.

\section{Conclusion and some further problems}

\begin{itemize} %\vspace{-8pt}
\addtolength{\itemsep}{-8pt}
\item
Quantum field theory is not yet completely
investigated for large multiplicities
of identical particles ($n\sim1/\l$).

\item
Both conventional
perturbation theory and the $1/N$ expansion break down at $n\sim1/\l$
or $n\sim N$.

\item
While for $n\sim1/\l$ or $n\sim N$ the problem becomes semiclassical,
there are subtleties in application of semiclassical technique.
Alternative methods could be useful.

\item
The mechanism of unitarity restoration is not yet completely work out
(while there is no problems with unitarity at large $N$). An exponentiation
of the corrections in $n^2\l$ or $n^2/N$ looks promising.

\item
Cross-sections of multiparticle production
may become large at very high energies $E\sim m/\l$ while smaller
than the unitary bound.

\item
The nullification survive at large $N$ to all loops (factorization is
crucial for the reduction to a quantum mechanical problem
at the tree and one-loop levels or at large $N$).

\item
Which dynamical symmetry is behind the nullification is not yet found out
and higher conservation laws which restrict multiparticle production are not
yet constructed.

\item
Drastic simplifications occur for the given kinematics
(in contrast to the case when all momenta are off mass shell).
The threshold amplitudes are described by an integrable lower dimensional
problem.

\item
One might try to solve more complicated $D=4$ models imposing this
kinematics.

\item
The developed technique might be applicable to some other problems.

\item
The nullification could have some phenomenological consequences for the
standard model restricting the ratio of masses.

\item
An extension to nonvanishing spatial momenta would be very interesting,
in particular, for
matching~\cite{CM94} the threshold behavior of multiparticle amplitudes
with the standard multiperipheral picture of
high energy collisions.

\vspace{-6pt} \end{itemize}

%\eop

\end{document}